\documentclass[twocolumn,showpacs,preprintnumbers,amsmath,amssymb]{revtex4}

\usepackage{graphicx}
\usepackage{dcolumn}
\usepackage{bm}
\usepackage[a4paper]{hyperref}

\begin{document}
\preprint{manuscript}

\title{Tunable effective g-factor in InAs nanowire quantum dots}

\author{M. T. Bj\"{o}rk, A. Fuhrer, A. E. Hansen, M. W. Larsson, L. E. Jensen, L. Samuelson}
\affiliation{Solid State Physics/Nanometer Structure
 Consortium, Lund University, P.O. Box 118 Lund, Sweden}
\email{andreas.fuhrer@ftf.lth.se}

\date{\today}

\begin{abstract}
We report tunneling spectroscopy measurements of the Zeeman spin splitting in InAs few-electron quantum dots. The dots are formed between two InP barriers in InAs nanowires with a wurtzite crystal structure which are grown using chemical beam epitaxy. The values of the electron $g$-factors of the first few electrons entering the dot are found to strongly depend on dot size. They range from close to the InAs bulk value in large dots $|g^*|=13$ down to $|g^*|=2.3$ for the smallest dots. 
\end{abstract}

\pacs{ 73.23.Hk, 73.63.-b, 73.63.Kv, 71.70.Ej}
\maketitle

The spin of an electron in a quantum dot (QD) is one of the candidates
for a  scaleable quantum bit, the fundamental unit in quantum computation
and quantum communication schemes~\cite{98loss}.
Experimental realizations are on the one hand pursued using top-down approaches. This usually involves lateral gate electrodes electrostatically confining few or a single electron in a two dimensional electron gas close to the surface of a Ga(Al)As based heterostructure~\cite{03hanson}. Such systems offer good tunability and controlled coupling of multiple spins has been demonstrated~\cite{04craig}. On the other hand, bottom up systems such as self assembled QDs~\cite{04kroutvar} and carbon nanotubes~\cite{05biercuk} are expected to scale more easily. Semiconductor nanowires have emerged as a promising bottom-up fabricated system for electronic and optical device applications~\cite{03samuelson}.  We have recently demonstrated the creation of few-electron QDs using InAs nanowire heterostructures~\cite{04bjork} with two InP barriers.
In the following we set out to investigate the spin properties of the first few orbital levels of these QDs. 

\begin{figure}[t!]
\centering
\includegraphics[width=8cm]{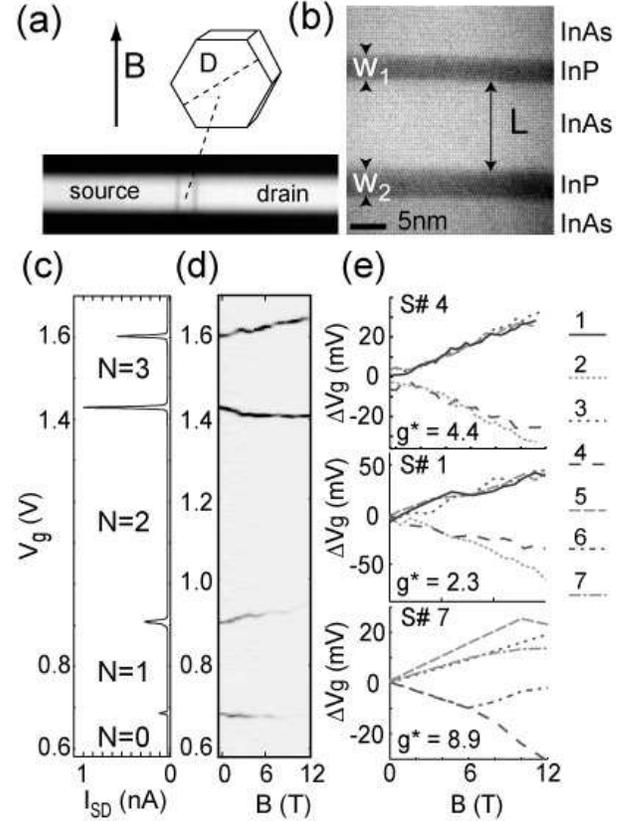}
\caption{\label{fig:1}\small  (a) STEM image of a nanowire together with a schematic of the InAs QD of diameter $D$.  (b) High resolution STEM image of a QD from the same growth as S\#4 and S\#6. The dot length $L=$12\;nm is deduced from the length of the InAs segment (bright) between two InP barriers (dark) having widths $w_1$ and $w_2$. (c) Coulomb blockade oscillations in $I_{SD}$ as a function of $V_g$ at $B=$0\;T and a bias $V_{SD}=50\;\mu$V for S\#4. Each peak corresponds to adding a single electron to the dot. (d) Dot conductance as a function of magnetic field and gate voltage. The peaks shift as a function of $B$ because of the Zeeman effect and small $B$ induced orbital shifts. (e) Peak separation $\Delta V_G$ as a function of $B$ showing the Zeeman splitting for the first electrons added to the dot.}\end{figure}
We utilize transport spectroscopy to measure the Zeeman splitting of the energy levels as a function of magnetic field and thereby determine the effective electron $g$-factor ($g^*$).
The g-factor of bulk InAs, which crystalizes in the zinc-blende (ZB) structure, has been found to be $g^*=-14.7$~\cite{82landolt}. However, InAs nanowires can exhibit both zinc-blende and wurtzite (WZ) type crystal structure\cite{92koguchi} depending on diameter and growth conditions and so far very little is known about band parameters in WZ InAs. 
In low-dimensional semiconductor heterostructures the $g$ factor depends critically on system size and dimensionality~\cite{98kiselev}. We show that varying the size of our nanowire dots allows us to tune $g^*$ from a value close to the InAs bulk value down to  $|g^*|=2.3\pm0.3$. The possibility to have multiple dots along a nanowire, each with a different g-factor, makes such systems interesting candidates for realizations of individually addressable spin qubits.

Using chemical beam epitaxy InAs nanowires containing QDs were grown catalytically from Au nanoparticles deposited on a $<$111$>$B InAs substrate~\cite{04jensen,02Abjork}. The nanowires typically grow perpendicular to the substrate and high resolution scanning transmission electron microscope (STEM) images indicate that most of them show a nearly perfect WZ lattice with very few twinning boundaries or stacking faults.
 The QDs were defined between two InP tunnel barriers within the InAs nanowires. This yields a confining potential for electrons set by the $600$~meV conduction band offset between InAs and InP in the growth direction~\cite{02Abjork}. In the lateral direction the side facets of the wire form a hexagonal cross section with presumably hard wall conditions [see Fig.~\ref{fig:1}(a)]. 
The wires studied here had diameters $D$ in the range of $50-70$\;nm. 
Five growths were made with wires containing QDs of different lengths. The dot length $L$ and the thicknesses $w_1$ and $w_2$ of the InP tunnel barriers were measured for each growth using STEM images such as the one shown in Fig.~\ref{fig:1}(b). 
A small variation in $D$ within the same growth (given by the size distribution of the gold catalyst particles) leads to slightly different growth rates\cite{04jensen}. The STEM images therefore allow us to establish a relation between $D$ and $L$,$w_1$,$w_2$ for each growth. We then measure the diameter $D$ of each nanowire in the actual device with a scanning electron microscope, which allows us to give estimates for all the size parameters accurate to within about 15\% (see Table~\ref{tab:1}). 

For transport measurements the samples were prepared by depositing wires on a degenerately doped Si substrate with a 100\;nm SiO$_{2}$ layer acting as the gate dielectric for the Si back gate. After locating individual wires, Ni/Au source and drain contacts were defined using electron beam lithography and metal evaporation. Device S\#8 was fabricated as a reference from a sixth growth of homogenous InAs nanowires without QDs. Here, local gates were used to define large QDs ($L\approx 200$\;nm)\cite{05fasth}. 

Electrical measurements were performed under symmetric bias conditions in a He-3 cryostat with a base temperature of $250$\;mK. The heterostructure dot lengths range from $L=8 - 20$\;nm which is much smaller than $D$. This leads to strong quantum confinement along the wire, so that only the lowest quantum state is occupied in the growth direction. A magnetic field $B$ is applied perpendicular to the wire. 

\begin{table}
\centering
\parbox{8.5cm}{
\begin{tabular}[c]{|c|c|c|c|c|c|c|}
\hline
S~\# & $L$ (nm) & $D$ (nm) & $w_{1}$ (nm) & $w_{2}$ (nm) & $G  (e^{2}/h)$ & $\vert g^{*}\vert$ \\
\hline
1 & 8 & 70 & 3 & 3.25 & 2 & $2.3\pm0.3$ \\
2 & 10 & 55 & 3.5 & 3.7 & 0.2 & $3.5\pm0.5$ \\
3 & 12 & 55 & 12 & 12 & 4$e^{-4}$ & $4.0\pm0.5$ \\
4 & 12 & 55 & 3 & 5 & 0.1 & $4.4\pm0.5$ \\
5 & 13 & 51 & 1.2 & 2.2 & 1 & $5.8\pm0.5$ \\
6 & 14 & 52 & 6 & 8 & 5$e^{-6}$ & $6.0\pm0.5$ \\
7 & 20 & 53 & 6 & 6 & 0.01 & $8.9\pm1.0$Ê\\
8 & 270 & 65 & - & -  & - & $13.0\pm2.5$Ê\\
\hline
\end{tabular}}
\caption{\label{tab:1}\small  Summary of important parameters for the different devices measured, including the measured effective g-factors averaged over data extracted from peak position and excited state spectroscopy.}
\end{table}
Figure~\ref{fig:1}(c) shows Coulomb blockade (CB) oscillations in the current $I_{SD}$ through the QD in S\#4 with $L=$12\;nm. At zero applied gate voltage the QD is empty ($N=0$). For $V_g = 686$\;mV the lowest orbital level of the QD aligns with the chemical potential in the source and drain contacts giving rise to a peak in $I_{SD}$. Upon adding more electrons the Coulomb peak positions on the $V_g$ axis directly reveal the addition spectrum of the QD for which a clear shell structure is observed~\cite{04bjork}. Up to at least the 10th electron the shell structure is solely determined by the lateral confinement. For higher electron numbers the second subband gets occupied for the largest heterostructure dot with $L=20$\;nm. The enhanced peak spacing between the second and third peak corresponds to a filled first shell i.e. the lowest orbital containing both a spin up and a spin down electron. 
The spin degeneracy is lifted at finite $B$ due to the Zeeman splitting $\Delta E_z=g^{*}\mu_B B$ where $\mu_B$ is the Bohr magneton. The Zeeman shift of the energy levels manifests itself directly as a change in the CB-peak positions~\cite{01kouwenhoven}. Figure~\ref{fig:1}(d) shows the CB peaks as a function of $B$ where peaks reflecting the charging of the same orbital level with spin up and spin down electrons move apart. Such spin pairing, indicating alternating spin filling for the first few orbital levels, is observed for all our dots except the two largest ones (S\#7 and S\#8) in strong magnetic fields. In addition to the Zeeman effect the peak positions are also shifted due to a small diamagnetic shift. Such contributions can be eliminated by taking the difference between two peak positions for which the same orbital is filled i.e. spin paired peaks neighbouring a gap with odd electron number. The resulting peak spacing $\Delta V_g$ is proportional to the addition energy $U(N)$ which can be separated into three parts $U(N)=E_{int} + \Delta E(N) + \Delta E_z$. Here $E_{int}$Ê is the interaction contribution, which within the constant interaction model\cite{01kouwenhoven} is simply  the charging energy $E_{int}=e^2/C_{dot}$, $\Delta E(N)$ is the single particle level spacing and $\Delta E_z$ is the Zeeman splitting. The capacitive lever arm between $V_g$ and dot energy can be extracted for each CB peak from measurements of Coulomb diamonds and is found to be in the range of $\alpha=C_{dot}/C_g=0.05-0.25$ depending mainly on dot size. The effective $g$-factor is now given from the slope of the curves obtained by calculating $\Delta V_g(B)$ as shown in Fig.~\ref{fig:1}(e) for three different dots. The topmost graph gives the first five peak spacings for dot S\#4, where the spin pairs are given by the positive slopes corresponding to odd numbered gaps [see legend in Fig.\ref{fig:1}(e)]. Fits to the measured data for these spin pairs yield $|g^*|=4.4\pm0.5$. The same type of analysis was done for all eight samples giving g-factors between $|g^*|=2.3\pm0.3$ for a dot with $L=8$\;nm (S\#1) and  $|g^*|=8.9\pm1.0$ for a dot with $L=20$\;nm (S\#7). For the latter case the large spin splitting leads to Zeeman induced crossings, which explains the deviations from the linear behaviour of $\Delta V_g$. In addition, we were not able to measure the first two diamonds with sufficient accuracy in this sample, which is why the corresponding lines are missing in the data.

\begin{figure}
\centering
\includegraphics[width=7cm]{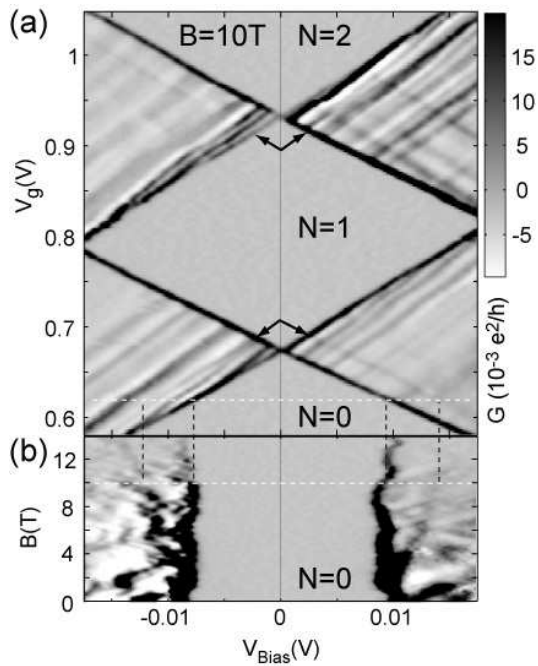}
\caption{\label{fig:2}\small  (a) Differential conductance as a function of gate voltage and applied bias for S\#3 at $B=10$\;T. The Zeeman split excited states are seen as lines running parallel to the diamond edges as indicated by the arrows.  (b) Differential conductance at a gate voltage of $620~mV$ as a function of bias and magnetic field where the linear Zeeman splitting is directly observed.}
\end{figure}
We also determine the Zeeman splitting directly from the excited state spectrum obtained in Coulomb diamond measurements. Fig.~\ref{fig:2}(a) shows the differential conductance $G$ as a function of $V_G$ and $V_{SD}$ in grayscale at $B=10$\;T again for S\#4. For the lowest open diamond the dot is empty ($N=0$). The borderlines of this diamond mark the situation where either the source or drain chemical potential is aligned to the first dot level being charged with a single spin up electron. The Zeeman splitting is then observed as an excited state line moving parallel to the borderlines but shifted to larger gate voltages\cite{03hanson}(see arrows). The same is true for excitations below and parallel to the borderlines of the $N=1$ to $N=2$ transition. The corresponding transitions are not observed within the $N=2$ diamond since the first shell can accommodate two spins only and the total spin of the dot is expected to be zero in that situation.
The lines outside the diamonds that have a different slope than the diamond edges and show negative differential conductance (white lines) are due to a modulation of transport through the dot by states in the quasi one-dimensional source and drain contacts. In order to make the difference between these lead states and excited state lines more clear, Fig.~\ref{fig:2}(b)  shows $G$ as a function of $B$ and source-drain bias for fixed $V_G=620$\;mV [dashed white horizontal line in Fig.~\ref{fig:2}(a)]. Here the linear Zeeman splitting is directly observed as a function of $B$. The lead states appear as a fast modulation of the conductance and to a lesser degree also influence the position of the diamond borderlines. The splitting at $B=10$\;T is outlined by vertical dashed lines. Measuring the splitting as a function of $B$ in Fig.~\ref{fig:2}(b) we deduce $|g^*|=4.3\pm0.5$ in good agreement with $|g^*|=4.4\pm0.5$ found above from peak separations.

\begin{figure}
\centering
\includegraphics[width=7.5cm]{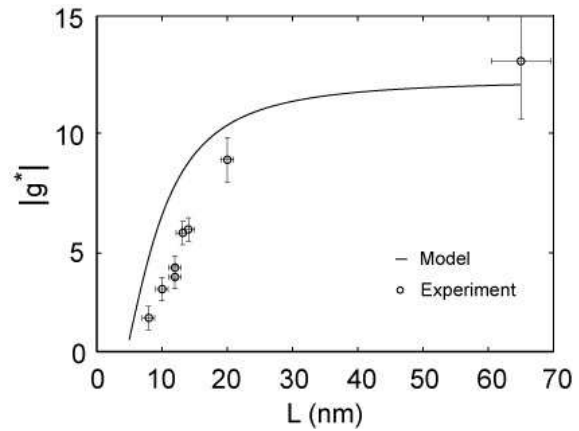}
\caption{\label{fig:3}\small  Dependence of dot length on the effective $g$-factor. The top curve is the theoretical values obtained from Eq.~(\ref{eq:gfactor}) and hard wall confinement. }
\end{figure}

The effective $g$-factors for $8$ dots of different lengths were deduced in the same way as above and are plotted as a function of $L$ in Fig.~\ref{fig:3}. Here, $L$ denotes the extent of the dot in the strongest confinement direction. For S\#8, which was the dot defined using local gate electrodes, we therefore replaced $L$ by the diameter $D$ of the nanowire since the separation of the gate electrodes was more than 200\;nm. We find that in this case $g^*$ approaches nearly the bulk value found for ZB InAs. For the heterostructure dots with small $L$, $g^*$ is strongly suppressed. 

Generally, $g$-factors in confined semiconductor systems have been found to differ significantly from the corresponding bulk values~\cite{98kiselev,94kowalski,88dobers,03nitta}. In measurements on InAs-based two-dimensional electron systems, values from $g^*=-1$ to $g^*=-13$ have been found experimentally depending on magnetic field direction~\cite{00brosig}, quantum well width~\cite{94kowalski} and barrier composition~\cite{94kowalski,03nitta} and gate tunability of $g^{*}$ has been demonstrated\cite{03nitta,01salis}. 
Investigations on self-assembled InAs QDs using magneto-tunneling and capacitance spectroscopy have revealed anisotropic $g$-factors with values from +0.5 to +1.6~\cite{98thornton,02medeiros}. 
For WZ crystals both effective mass and g-factor are expected to be anisotropic and theoretical expression can be derived within $k\cdot p$-theory\cite{77hermann,01bayerl}. However, to the best of our knowledge, band parameters for WZ InAs are not known. We therefore, start the discussion of our results with a comparison to a simple expression derived for cubic III/V-compounds\cite{77hermann}. 
 \begin{equation}
   \frac{g^{*}}{g}= 1- \frac{P^2}{3}\frac{\Delta_{so}}{E_{g}(E_{g}+\Delta_{so})}
   \label{eq:gfactor}
\end{equation}

Here $g^*$ depends on the bandgap $E_{g}$ of the semiconductor material and on the spin-orbit splitting $\Delta_{so}$ in the valence band. $P^2$ is a band parameter related to the interband momentum matrix elements. The suppression of  $g^{*}$ can be understood qualitatively in a simple picture in which the dot confinement leads to an enhanced gap between electron and hole states. 
We therefore assume that $E_g=E_g^{bulk}+E_1$ where $E_1$ is the confinement energy for the lowest subband of the dot calculated using a hard wall confinement potential with width $L$.
For WZ semiconductors the gap is typically slightly larger than for the same ZB material\cite{94murayama}. Accordingly, we use $E_g^{bulk}=0.460$\;eV a value which is 10\% larger than the ZB value. Beyond that, we rely on the band parameters for ZB InAs which we expect to be a good first approximation\cite{94murayama,01bayerl}. 
In Fig.~\ref{fig:3} the solid line shows a curve derived from Eq.~(\ref{eq:gfactor}) with $P^2=21.5$\;eV, $\Delta_{so}=0.39$\;eV and a constant effective electron mass of $0.026$\cite{01vurgaftman}. 
While this model reproduces the general trend of the measured data points and fits reasonably well for the larger dots, there is a much stronger suppression of $g^{*}$ with decreasing $L$ in the experiment. 

We have considered the reduction in $g^{*}$ for small dot sizes to be due to penetration of the wavefunction into the InP barriers which have a bulk $g$-factor of $+1.4$\cite{82landolt}. However, in the experiment we observe no measurable change in $g^{*}$ with electron number. For large $B$ the overall conductance is observed to strongly decrease due to a stronger magnetic confinement which results in weaker coupling but no observable change in $g^{*}$. 
In Table~\ref{tab:1} we also give values for the Coulomb peak conductance $G$ at $B=0$\;T and small bias (average over first five peaks) which is clearly correlated to the barrier thickness. This demonstrates tunability of the coupling while keeping $L$ fixed. We do not find, however, that this is correlated with a change in $g^{*}$ (see e.g. S\#3 and S\#4). This leads us to conclude that wavefunction overlap with the barrier material is unlikely to be the cause of the more strongly reduced values of $g^{*}$ for small $L$ in the experiment. A more accurate model is necessary, taking into account both the confinement and the WZ crystal structure in order to explain the observed data.

In conclusion we have measured the effective electron $g$-factor of InAs QDs defined by InP double barriers in semiconductor nanowires with a WZ crystal structure as a function of dot size. We observe a strong suppression of $g^*$ for small dot length $L$ down to $|g^*|=2.3$ for S\#1 ($L=8$\;nm). While this trend can be understood qualitatively due to an increase of the confinement energy a good theoretical model for WZ InAs is missing. We hope that our experiments will spur the development of more quantitative models for these type of InAs QDs.
It was further shown that changing the dot length $L$ allows us to design QDs along a nanowire with different specific spin splittings in a constant magnetic field. Future development of graded heterostructures is expected to allow individual gate tunable spin splittings\cite{01salis,05fasth} in a series of dots along a nanowire, which makes nanowire QDs containing a single electron spin interesting systems for the realization of qubits. 

\begin{acknowledgments}
This work was supported by the Swedish Foundation for Strategic 
Research (SSF), the Swedish Research Council (VR), the Knut and Alice Wallenberg Foundation (KAW), the Office of Naval Research
(ONR), and the Swiss Science Foundation (Schweizerischer Nationalfonds). Helpful discussions with C. Thelander and S. Ulloa are gratefully acknowledged.
\end{acknowledgments}

\bibliographystyle{apsrev}

\end{document}